\documentstyle[twoside,npb]{article}
\begin{document}
\newcommand{\be}{\begin{equation}}
\newcommand{\ee}{\end{equation}}
\newcommand{\al}{\alpha}
\newcommand{\bt}{\beta}
\newcommand{\lm}{\lambda}
\newcommand{\bea}{\begin{eqnarray}}
\newcommand{\eea}{\end{eqnarray}}
\newcommand{\gm}{\gamma}
\newcommand{\Gm}{\Gamma}
\newcommand{\dl}{\delta}
\newcommand{\Dl}{\Delta}
\newcommand{\ep}{\epsilon}
\newcommand{\kp}{\kappa}
\newcommand{\Lm}{\Lambda}
\newcommand{\om}{\omega}
\newcommand{\pa}{\partial}
\newcommand{\dd}{\mbox{d}}
\newcommand{\MS}{\mbox{MS}}
\newcommand{\nn}{\nonumber}
\newcommand{\uk}{\underline{k}}
\newcommand{\ice}[1]{\relax}

\title{%
\vspace*{-40mm}
\begin{flushright}
{\normalsize  TTP/00--11\\[1mm]
hep-ph/0005301    \\[-1mm]
May 2000 }    \\[30mm]
 \end{flushright}
\vspace*{-20mm}
Subleading Sudakov Logarithms in Electroweak Processes
\thanks{Talk presented by J.H.K\"uhn 
}%
}
\author{%
J.H.~K\"uhn
\address{Institut f{\"u}r Theoretische Teilchenphysik,
Universit{\"a}t Karlsruhe, 76128 Karlsruhe, Germany},
A.A.~Penin\thanks{On leave from  Institute for Nuclear
Research of Russian Academy of Sciences, 117312 Moscow, Russia},
\address{II. Institut f{\"u}r Theoretische Physik,
Universit{\"a}t Hamburg, 22761 Hamburg, Germany} and
V.A.~Smirnov
\address{Nuclear Physics Institute of Moscow State University,
119899 Moscow, Russia}
}
\begin{abstract}
Recent results for the asymptotic behavior of 
fermion scattering amplitudes in the Sudakov
limit are presented including next-to-leading logarithmic corrections.
These are used for  the analysis of the
dominant electroweak corrections to the fermion-antifermion pair
production in $e^+e^-$ annihilation at high energy.
\end{abstract}

\thispagestyle{empty}

\maketitle

\section{Introduction}
Four fermion processes are generally considered as
benchmark processes at high energy colliders, with electron
positron annihilation into muon or quark pairs at LEP and the 
Drell Yan
process at hadron colliders as characteristic examples. Within the
presently accessible energy region, typically up to 200~GeV, radiative
corrections are dominated by the shift in the $W$ and $Z$ masses as
parametrized by the $\rho$ parameter and by the running of the 
coupling
constant. Vertex corrections and box diagrams involving gauge 
bosons are
generally of minor importance. In the TeV region, accessible at future
colliders like the LHC or TESLA, this picture changes drastically. 
A new
class of effects starts to become relevant and rapidly dominant which
are generally denoted as double logarithmic corrections and which were
first observed by Sudakov \cite{Sud}
in the context of quantum electrodynamics for
reactions with a tight cut on the radiated energy of the photons.
For electroweak interactions large negative corrections arise from the
exchange of gauge bosons which remain uncompensated if one restricts 
the
analysis to exclusive final states, consisting e.\ g.\ of a fermion
antifermion pair only.
The discussion of double logarithmic corrections is fairly
straightforward for a theory with massive gauge bosons only.
An important complication arises from the presence of massless photons
in the final state. Events with soft and hard photon radiation are
normally included in the sample -- whence a
``semiinclusive'' definition of the cross section is closest to the
the actual experimental analysis.

One loop corrections to the four fermion process are available since
long (see e.\ g.\ \cite{Hol}). In this case the separation of photonic
and weak corrections is still possible and, employing  axial 
gauge, the
leading logarithms can be trivially attributed to fermion self energy
diagrams with virtual $W$ or $Z$ boson emission \cite{KuhPen}.

Results for the leading logarithmic corrections in higher orders have
been presented in \cite{KuhPen,CiaCom,Fad,KPS} (for a more recent
discussion see also
\cite{Melles,Been}). The first three results disagree, a 
consequence of
different requirements on the inclusion or exclusion of hard
photon emmission and a different treatment of the virtual corrections.
In \cite{KPS} it was demonstrated that the approach developed in
\cite{KuhPen}, where axial gauge was adopted and
both real and virtual photon emmission neglected, is
numerically nearly identical to the results of \cite{Fad} where 
virtual
photon corrections and real photon emmission are included --- a
consequence of the smallness of the weak mixing angle, $\sin^2\theta_W
=0.23$. Furthermore, an analysis of subleading logarithms for the form
factor, for the four fermion scattering in a spontaneously broken
$SU(2)$ gauge theory and, last not least, for the Standard Model with
$W$, $Z$ and the massless photon was performed in \cite{KPS}.

Before presenting a brief review of the techniques and some results 
from
\cite{KPS} let us emphasize the importance of subleading corrections.
In
the energy region relevant at LHC or TESLA, i.e.\ around 1 to 2~TEV,
this is already evident from the one loop vertex correction.
In the timelike region
\be
|F_B + \delta F|^2 \approx F_B(1+
2\frac{\alpha_W}{\pi} \rho(s/M^2)),
\ee
\[
\rho(u)=
-\frac14\ln^2 u  + \frac 34\ln u - \frac{\pi^2}{12} + \frac 78 .
\]
For characteristic values of the coupling, energy and mass of
$\alpha_W/\pi = 10^{-2}$, $C_F=3/4$ and $s/M^2=10^2$ one finds
\[
2\frac{\alpha_W}{\pi} \rho =1.5 \times 10^{-2} 
( - 5.30 + 3.45  - 1.70)=
0.053.
\]
Large compensations between leading and subleading terms are observed
and, in fact, this pattern will reappear for the Standard Model as
discussed below.

The treatment of \cite{KPS} is based on evolution equations that 
govern
the dynamics of the amplitudes in the Sudakov limit
as obtained in \cite{Col,Sen1,Sen2,Kor1,Kor2}.
In \cite{KPS}
this approach was applied to the
next-to-leading analysis of the  Abelian form factor
and the four fermion amplitude in the $SU(N)$ gauge theory.
Functions that enter the evolution equations
in the  next-to-leading logarithmic approximation
were evaluated by using, as an
input, asymptotic expansions of one-loop diagrams.
The solution of these equations lead to a summation of the leading
and subleading Sudakov logarithms.
The  expansion of one-loop diagrams through the so-called
generalized strategy of regions \cite{BS} (see also \cite{SR})
identifies in a systematic way the nature
of various contributions and the origin of logarithms.
This strategy is based on expanding
integrands of Feynman integrals in typical regions and
extending the integration domains to the whole space of
the loop momenta so that a crucial difference with respect to
the standard approach \cite{Col,Sen1,Sen2,Kor1,Kor2} is the absence
of cut-offs that specify the regions in individual terms of
the expansions.

\section{The  Abelian form factor in the Sudakov limit}

Let us first analyse
the (vector) form factor  which determines the  amplitude of the
fermion scattering in the external Abelian field. In Born
approximation
\be
F_B=\bar\psi(p_2)\gm_\mu\psi(p_1)\; ,
\ee

We consider the limit
$s=(p_1-p_2)^2 \to-\infty$ with
on-shell massless fermions, $p_1^2=p_2^2=0$, and
gauge bosons with a small non-zero mass $M^2\ll -s$.
For convenience
$p_{1,2} = (Q/2,0,0,\mp Q/2)$ so that  $2 p_1 p_2 = Q^2=-s$.

The asymptotic behaviour
can be found by solving the corresponding evolution equation
 \cite{Col}
\be
{\partial\over\partial\ln{Q^2}}F=\mbox{\hskip4.6cm}
\label{eei}
\ee
\[\mbox{\hskip.4cm}
\Bigl[\int_{M^2}^{Q^2}{\dd x\over x}\gm(\al(x))+\zeta(\al(Q^2))
+\xi(\al(M^2)) \Bigr] F \; .
\]
For the non-Abelian gauge theory, this  equation was first derived
in \cite{Sen1} by factorizing collinear logarithms in the
axial gauge.
Its solution is
\be
F=F_0(\al(M^2))\exp \Bigl\{\int_{M^2}^{Q^2}{\dd x\over x}
\mbox{\hskip2.5cm}
\ee
\[\times
\Bigl[\int_{M^2}^{x}{\dd x'\over x'}\gm(\al(x'))+\zeta(\al(x))
+\xi(\al(M^2))\Bigr]\Bigr\} \; .
\]
For a proper treatment of the
next-to-leading  logarithms one must
keep renormalization group corrections to the leading
logarithmic approximation
as well as single infrared and renormalization group
logarithms. In this approximation
\be
F=F_0(\al)\exp\Bigl[\int_{M^2}^{Q^2}{\dd x\over x}
\int_{M^2}^{x}{\dd x'\over x'}\gm(\al(x'))\mbox{\hskip.2cm}
\ee
\[
\mbox{\hskip2cm}
+(\zeta(\al) +\xi(\al))\ln{(Q^2/M^2)}\Bigr]
\]
The leading terms of
the functions $\gm$, $\zeta$  and $\xi$ are obtained from the one loop
analysis
and the one loop running of $\al$ in the argument of the
function $\gm$ should be taken into account.

In the covariant gauge, the self energy insertions
to the external fermion lines do not give $Q$-dependent
contributions.
The one loop calculation of the vertex correction gives
\be
F={\al\over 2\pi}C_F\left(-V_0+2V_1+2(1-2\ep)V_2-V_2'\right)F_B\, ,
\ee
where $C_F$ is the  quadratic Casimir operator of
the fundamental representation
and the functions $V_i$  are obtained from
\be
\int \frac{\dd^d k }{(k^2-2p_1k) (k^2-2p_2k) (k^2-M^2)} =
\mbox{\hskip0.5cm}
\label{ints}
\ee
\[
\mbox{\hskip4.3cm}
i\pi^{d/2}e^{-\gm_{\rm E}\ep} s^{-1} V_0  \; ,
\]
\[
\int \frac{\dd^d k \; k_\mu}{(k^2-2p_1k) (k^2-2p_2k) (k^2-M^2)} =
\mbox{\hskip1.0cm}
\]
\[
\mbox{\hskip2.5cm}
i\pi^{d/2}e^{-\gm_{\rm E}\ep} s^{-1} (p_1+p_2)_\mu V_1 \; ,
\]
\[
\int \frac{\dd^d k \; k_\mu k_\nu}{(k^2-2p_1k) (k^2-2p_2k) (k^2-M^2)}
= \mbox{\hskip1cm}
\]
\[
 \mbox{\hskip1cm}
i\pi^{d/2}e^{-\gm_{\rm E}\ep}
\left[g_{\mu\nu}V_2+{{p_1}_\mu{p_2}_\nu
+ (\mu\leftrightarrow\nu)\over s}V_2' \right]
\]

To expand these integrals in the limit $Q^2\gg  M^2$ we apply a 
generalized
strategy of regions formulated in \cite{BS} and
discussed using characteristic two-loop examples in~\cite{SR}:
\begin{itemize}
\item Consider various regions of the loop momenta and expand, in
every region, the integrand in Taylor series with respect to the
parameters that are there considered small;
\item Integrate the  expanded integrand
over the whole integration domain of the loop momenta;
\item Put to zero any scaleless integral.
\end{itemize}

The following ``typical'' regions arise in the Sudakov
limit \cite{Ste}:
\bea
\label{h}
\mbox{{\em hard} (h):} && k\sim Q
\nn \\
\label{1c}
\mbox{{\em 1-collinear} (1c):} && k_0 + k_3 \sim Q\nn \\
&& k_0 - k_3 \sim M^2/Q\, ,
\,\, k_{1,2} \sim M
\nn \\
\label{2c}
\mbox{{\em 2-collinear} (2c):} && k_0 - k_3 \sim Q\nn \\
&& k_0 + k_3 \sim M^2/Q\, ,
\,\, k_{1,2} \sim M 
\nn \\
\label{soft}
\mbox{{\em soft} (s):} && k\sim M \nn \\
\label{us}
\mbox{{\em ultrasoft} (us):} && k\sim M^2/Q\, .
\nn
\eea

Keeping the leading power in the expansion in the limit
$Q^2/M^2\to \infty$ one observes \cite{KPS} that the leading double
logarithm results from the hard region, whereas the single logarithm
receives contributions from the hard and collinear regions as well.
Soft regions do not contribute,
at least in the leading power.
Combining the vertex correction with the $Q^2$ independent fermion self
energies one arrives at a finite result.

 From the one-loop result  one derives
\be
\gm(\al)=-C_F{\al\over 2\pi} \;.
\ee
The total double logarithms originate
from the hard region. This determines
the scale of the coupling constant in the second order
logarithmic derivative of the form factor in $Q$.
At the same time we cannot distinguish, in the one loop approximation,
the contribution to the functions $\zeta$  and $\xi$
coming from the collinear region
because this region includes both $Q$ and $M$ scales.
For the sum of these functions we find
\be
\zeta(\al)+\xi(\al)=3C_F{\al\over 4\pi} \; .
\ee
Finally, in the NLO logarithmic approximation
\[
F = F_B\left(1-C_F{\al\over 2\pi}\left({7\over 2}+
{2\pi^2\over 3}\right)\right)
\exp \Bigl\{{C_F\over 2\pi}\Bigl[\mbox{\hskip3mm}
\] 
\be
\mbox{\hskip3mm}
-\int_{M^2}^{Q^2}{\dd x\over x}
\int_{M^2}^{x}{\dd x'\over x'}\al(x')
 +3\al\ln{(Q^2/M^2)}\Bigr]\Bigr\}
\label{fi}
\ee
in agreement with the result of \cite{Kor1,Kor2}.

\section{The four fermion amplitude}
We study  the limit of the fixed-angle scattering when
all the  invariant energy and momentum transfers
of the process are much larger than the
typical mass scale of internal particles 
$|s|\sim |t| \sim |u| \gg M^2$.
Besides the extra kinematical variable the analysis of the four
fermion amplitude is more complicated  by the presence of different
``color'' and Lorentz structures.
The Born amplitude, for example,  can be expanded in the
basis of color/chiral amplitudes
\be
A_{B}={ig^2\over s}A^\lm
={ig^2\over s}T_F\Bigl(-{1\over N}
\left(A_{LL}^d+A_{LR}^d\right)
\ee
\[
\mbox{\hskip3cm}
+A_{LL}^c+A_{LR}^c
+(L\leftrightarrow R)\Bigr) \;,
\]
where
\bea
A^\lm~~&=&
\bar\psi_2(p_2)t^a\gm_\mu\psi_1(p_1)
\bar\psi_4(p_4)t^a\gm_\mu\psi_3(p_3)\, , \nn \\
A_{LL}^d&=&
\bar{\psi_2}_L^i\gm_\mu{\psi_1}_L^i
\bar{\psi_4}_L^j\gm_\mu{\psi_3}_L^j \, , \\
A_{LR}^c&=&
\bar{\psi_2}_L^j\gm_\mu{\psi_1}_L^i
\bar{\psi_4}_R^i\gm_\mu{\psi_3}_R^j
\nn
\eea
and so on.
Here $t^a$ is the $SU(N)$ generator,
$p_1$, $p_3$ are incoming and $p_2$, $p_4$ outgoing momenta
so that $t=(p_1-p_4)^2$ and $u=(p_1+p_3)^2=-(s+t)$.
For the moment we consider a parity conserving theory. Hence
only two chiral amplitudes are independent,
for example, $LL$ and $LR$.
Similarly only two color amplitudes are independent,
for example, $A^\lm$ and $A^d$.

Let us first compute the one loop corrections, employing again the
strategy of regions. The total   contribution from vertex and box
diagrams in the logarithmic approximation
is independent from chirality and the same both for
the $LL$ and  $LR$ amplitudes:
\be
{ig^2(Q^2)\over s}{1\over 2}
\bigg[
\bigg\{-C_F L(s)+ \bigg(3C_F-C_A\ln\left({u\over s}\right)
\label{4fer}
\ee
\[
\mbox{\hskip2.5cm}
+2\bigg(C_F-{T_F\over N}\bigg)\ln\left({u\over t}\right)
\bigg) l(s)\bigg\} A^\lm
\]
\[
\mbox{\hskip2cm}
+\bigg\{2{C_FT_F\over N}\ln\left({u\over t}\right)
\ln\left({-s\over M^2}\right)\bigg\} A^d
\bigg]
\]
with
\[
L(s)={g^2\over 16\pi^2}\ln^2\left({-s\over M^2}\right)\,; \,\,\, 
l(s)={g^2\over 16\pi^2}\ln\left({-s\over M^2}\right)
\]
and the same both for
the $LL$ and  $LR$ amplitudes.

Now  the collinear  logarithms can be separated from the
total one-loop correction.
For each fermion-antifermion pair, they
form the  exponential factor
found in the previous section (eq.~(\ref{fi})).
This factor in addition incorporates the
renormalization group logarithms which are not absorbed by
changing the normalization scale of the gauge coupling.
The rest of the single logarithms in eq.~(\ref{4fer})
is of the soft nature.
Let us denote by   $\tilde A$   the amplitude with
the collinear logarithms  factored out.  It can be  represented 
as a vector
in the basis $A^\lm$, $A^d$ and satisfies the following evolution 
equation
\cite{Sen2,Bot}
\be
{\partial \over \partial \ln{Q^2}}\tilde A=
{\bf \chi}(\al(Q^2))\tilde A \; ,
\label{evol3}
\ee
where  ${\bf \chi}$ is the matrix of the ``soft''
anomalous dimensions.
 From eq.~(\ref{evol3}) we find the elements of this matrix to be,
in units of $\al/ (4\pi)$,
\bea
\chi_{\lm \lm} &=&
-2C_A\ln\left({u\over s}\right)+
4\left(C_F-{T_F\over N}\right)\ln\left({u\over t}\right)  \nn \\
\chi_{\lm d} &=&4{C_FT_F\over N}\ln\left({u\over t}\right)  \nn \\
\chi_{d \lm} &=& 4\ln\left({u\over t}\right) \label{mat} \\
\chi_{d d} &=& 0\; .  \nn
\eea
The solution of  eq.~(\ref{evol3}) reads
\be
\tilde A =
A^0_{1}(\al(M^2))\exp{\left[\int_{M^2}^{Q^2}{\dd x\over x}
\chi_1(\al(x))\right]}
\ee
\[
\mbox{\hskip0.5cm}
+A^0_{2}(\al(M^2))\exp{\left[\int_{M^2}^{Q^2}{\dd x\over
x}\chi_2(\al(x))\right]} \;, 
\]
where $\chi_i$ are eigenvalues of the ${\bf \chi}$ matrix and
$A^0_i$ are $Q$-independent vectors.

>From the asymptotic expansion of the box dia\-grams
one finds that only the hard parts
contribute to eq.~(\ref{evol3}). This fixes the scale of $\al$ in this
equation  to be $Q$.

In the Abelian case, there are no different color
amplitudes and there is only one anomalous dimension
\be
\chi=4\ln\left({u\over t}\right)\, .
\ee

\section{Sudakov logarithms in electroweak processes}
We are interested in the process
$f'\bar f'\rightarrow f\bar f$.
In the Born approximation, its amplitude is of
the following form \cite{KuhPen}
\be
A_{B}={ig^2\over s}\sum_{I,~J=L,~R}\left(T^3_{f'}T^3_{f}+
t^2_W{Y_{f'}Y_{f}\over 4}\right)A^{f'f}_{IJ}\; ,
\ee
where
\be
A^{f'f}_{IJ}=\bar f_I'\gamma_\mu f_I'
\bar f_J\gamma_\mu f_J \; ,
\ee

To analyze the  electroweak correction to the above process
we use the approximation with
the $W$ and $Z$ bosons of the same mass  $M$
and massless quarks and leptons.
The photon is  massless, and
the corresponding  infrared divergent contributions  should be
accompanied by  the real soft photon radiation
integrated to some resolution energy  $\omega_{res}$
to get an infrared safe cross
section independent on an auxiliary photon mass.
At the same time the massive gauge bosons are supposed
to be  detected as separate particles.
In practice, the  resolution energy
is much less than the $W$ ($Z$) boson mass so the   soft photon
emission is of the QED nature. This cancels the  infrared 
singularities
of the QED virtual correction. We should therefore separate the 
QED virtual
correction from the complete result computed with the photon
of some mass $\lm$ and then  evaluate the QED virtual corrections
together with the real soft photon radiation effects with 
vanishing $\lm$.
It is convenient to subtract the QED contribution
computed with the photon of the  mass $M$  from the obtained result
for the virtual corrections and then take the limit $\lm \to 0$ 
for the
sum of QED virtual and real photon contributions to the total 
amplitude.
In the language of the
approach of ref.~\cite{Fad}, this prescription means that we use the
auxiliary photon  mass $\lm$  as a variable of the evolution
equation below the scale $M$ and the  subtraction fixes a relevant
initial condition for this  differential equation.
This leads to a modification of the  collinear
factor
and the soft anomalous dimensions.

Thus we keep always a cutoff $\omega_{res} \ll M$ and do not display the
QED Sudakov factor arising from this suppression of photon radiation.
The  remaining ``electroweak''universal collinear factor for   each
fermion-antifermion  pair  becomes
\be
\exp\left[
-\left(T_f(T_f+1)+t_W^2{Y_f^2\over 4}-s_W^2Q_f^2\right)\right.
\mbox{\hskip1cm}
\label{col}
\ee
\[
\mbox{\hskip3cm}
\times\left(L(s)- 3l(s)\right)  \Bigr].
\]
The  soft anomalous dimension for $I$ and/or $J=R$
is Abelian and, in units of $g^2 / (16\pi^2)$, reads
\be
\chi=\left(t_W^2Y_{f'}Y_f-4s_W^2Q_{f'}Q_f\right)
\ln\left({u\over t}\right)\;,
\ee
and the  matrix of the soft anomalous dimension for $I=J=L$ is
\newpage
\bea
\chi_{\lm \lm} &=&
-4\ln\left({u\over s}\right)\nn\\
&&+ \left(t_W^2Y_{f'}Y_f-4s_W^2Q_{f'}Q_f+2\right)
\ln\left({u\over t}\right)
\nn \\
\chi_{\lm d} &=&{3\over 4}\ln\left({u\over t}\right) \nn \\
\chi_{d \lm} &=& 4\ln\left({u\over t}\right) \\
\chi_{d d} &=& \left(t_W^2Y_{f'}Y_f-4s_W^2Q_{f'}Q_f\right)
\ln\left({u\over t}\right)\; . \nn
\eea
The one-loop leading and subleading  logarithms
can be directly obtained
from eq.~(\ref{4fer}).
The two-loop  leading (infrared)
logarithms  are determined by the
second order term  of the
expansion of the double  (soft$\times$collinear) logarithmic
part of the collinear  factors~(\ref{col}).
The two-loop next-to-leading   logarithms
are generated by the  interference between the first order terms
of the expansion of the
double (soft$\times$collinear) and single (soft$+$collinear$+
$re\-nor\-ma\-li\-zation group) logarithmic  exponents
and can also be found in \cite{KPS}.
With the expression for the  chiral amplitudes
at hand, we can compute the leading and subleading logarithmic 
corrections
to the basic observables for $e^+e^-\rightarrow f \bar f$.
Let us, for example, consider  the total cross sections
of the quark-antiquark/$\mu^+\mu^-$ production in
the $e^+e^-$ annihilation.
In the two loop approximation, the  leading and next-to-leading
Sudakov  corrections to the cross sections read
\[
\sigma/\sigma_B(e^+e^-\rightarrow Q\bar Q)
=1+\phantom{1}5.30l(s)-1.66L(s)
\hspace{0.5cm}
\]
\[
\hspace{3cm}
-12.84l(s)L(s)+1.92L^2(s) \; ,
\]
\[
\sigma/\sigma_B(e^+e^-\rightarrow q\bar q)
=1+20.54l(s)-2.17L(s)
\hspace{0.5cm}
\]
\[
\hspace{3cm}
-53.72l(s)L(s)+2.79L^2(s)\; ,
\]
\[
\sigma/\sigma_B(e^+e^-\rightarrow \mu^+\mu^-)
=1+10.09l(s)-1.39L(s)
\hspace{0.3cm}
\]
\be
\hspace{3cm}
-21.66l(s)L(s)+1.41L^2(s)\; ,
\label{sig}
\ee
where $Q=u,c,t$,  $q=d,s,b$.
Numerically, $L(s)=0.07$  $(0.11)$ and $l(s)=0.014$  $(0.017)$
respectively for $\sqrt{s}=1$~TeV and $2$~TeV.

Clearly, for energies at 1 and 2 TeV
the two loop corrections are huge and amount up respectively
to $5\%$ and $7\%$.
There is a cancellation between the leading and
subleading logarithms and for the above energy interval the
subleading contribution even exceeds the  leading one.
The higher order leading and next-to-leading
corrections however do not exceed $1\%$ level.
They can be in principle resummed using the formulae given above.
The leading and subleading corrections to the left right and forward
backward asymmetries \cite{KPS} are typically smaller.

Our result for the one loop  double logarithmic
contribution is in agreement with \cite{CiaCom}. However
the result for the   one loop single  infrared logarithmic
contribution differs from \cite{Bec}. The reason is that, 
in \cite{Bec},
only the diagrams with heavy virtual bosons have been taken into
account.  There is an infrared safe contribution of the diagram with
the virtual massless photon where the heavy boson
mass serves as an infrared regulator that should be taken into
account to get a complete (exponential) result.
In one-loop approximation,
this contribution comes from the box diagrams with
the photon and $Z$ boson running inside the loop \cite{KS}.

Our result for the two-loop  double logarithmic
contribution is in agreement with \cite{Fad}.
On the other hand, the coefficients in front of the two-loop
leading logarithms in eq.~(\ref{sig}) with a few percent
accuracy coincide  with the result of \cite{KuhPen} where
the photon contributions
were not considered. This is related to the fact
that  the virtual photon contribution not included to the result
of  \cite{KuhPen} is suppressed by a small factor $s_W^2$.
\\[3mm]
{\bf Acknowledgments}\\[3mm]
The work by V.S. was supported by the Volkswagen Foundation, contract
No.~I/73611. The work by J.K. and A.P. was supported by the Volkswagen
Foundation, by BMBF under grant BMBF-057KA92P and by 
DFG-Forschergruppe
``Quantenfeldtheorie, Computeralgebra und Monte-Carlo-Simulationen''
(DFG Contract KU502/6--1).

\end{document}